\input epsf
\documentstyle[prl,twocolumn,aps]{revtex}
\begin{document}
\draft
\title{
Low frequency spectroscopy of the correlated metallic system 
$Ca_xSr_{1-x}VO_3$.
}

\author{M. J. Rozenberg$^1$, I. H. Inoue$^2$, H. Makino$^3$,
F. Iga$^4$ and Y. Nishihara$^4$}
\address{1. LPT - 
Ecole Normale Sup\'{e}rieure, 24 rue Lhomond, 75005 Paris, France.\\
2. PRESTO Research Development Corporation of Japan,
and Electrotechnical Laboratory, Tsukuba 305, Japan.\\
3. Institute of Applied Physics, University of Tsukuba, Tsukuba 305,
Japan.\\
4. Electrotechnical Laboratory, Tsukuba 305, Japan.}

\maketitle

\begin{abstract}

We study the 
photoemission and optical conductivity response
of the strongly correlated metallic system $Ca_xSr_{1-x}VO_3$.
We find that the basic features of the 
transfer of spectral weight in photoemission experiments
and the unusual lineshape of the optical response
can be understood by modeling 
the system with a one band Hubbard model close to the 
Mott-Hubbard transition. We present a detailed
comparison between the low frequency experimental data and the corresponding 
theoretical predictions obtained within the LISA method that is
exact in the limit of large lattice connectivity.
\end{abstract} 

The understanding of the redistributions of the spectral weight
that occurs in strongly correlated systems is 
a fascinating problem of condensed matter physics \cite{zsa,tsuda,gunn}.

Most 
transition-metal oxides have narrow conduction
bands originated in partially filled $3d-$orbitals where correlation effects 
are significantly relevant. One of the classical examples is
the $V_2O_3$ compound and its derivatives which exhibit a Mott transition, 
that is, a metal-insulator transition (MIT) driven by electronic correlations 
in the absence of magnetic ordering \cite{mcwhan,rkk}. 
In a recent paper Fujimori {\it et al.} \cite{fuji}
examined the photoemission spectra of several
transition metal oxydes demonstrating the dramatic redistribution
of spectral weight as consequence of electronic correlations in those
compounds.
Here, we consider another correlated system, $Ca_xSr_{1-x}VO_3$, 
which displays a notable transfer of weight as a function of $x$ in its
photoemission spectra \cite{inoue,morikawa}, and very unusual features 
in the distribution of
its low frequency optical response.
This system is particularly interesting as it allows for {\it a
systematic study} of the evolution of the spectral lineshapes with $x$.
We shall argue that the basic features observed in a recent photoemission
study \cite{inoue,morikawa} and 
in the new optical conductivity results that are
reported here,
can be explained in terms of the solution of
a single band Hubbard model close to the MIT. 
We shall present a detailed comparison between the experimental data and
the theoretical predictions of the model treated within the Local Impurity
Selfconsistent Approximation (LISA)
which is exact in the limit of large 
connectivity, or equivalently, large dimensionality \cite{vollmetz,review,gk}.

The $Ca_xSr_{1-x}VO_3$ compound has a perovskite-type structure where 
the nominal valency of vanadium ions is $4+$, which leaves only
one $3d-$electron per vanadium ion in the conduction band. 
Thus, in a first approximation,  we shall
assume that the system can be modeled by 
a half-filled single band Hubbard hamiltonian. 
The hope is that this simple model may capture the basic features
of the low energy physics that is experimentally observed.

The hamiltonian of the Hubbard model reads,
\begin{equation}
H = -\sum_{<i,j>} (t_{ij}+\mu) c_{i\sigma}^{\dagger} c_{j\sigma}
 + \  U \sum_i  (n_{i \uparrow}-\frac{1}{2}) (n_{i\downarrow}-\frac{1}{2})
\label{HubHam}
\end{equation}
where summation over repeated spin indices is assumed.

The LISA method \cite{review} is a dynamical mean field theory
that is exact in the limit
$q \rightarrow \infty$, where
$q$ is the connectivity of the lattice.
Consequently, the hopping amplitude is rescaled as
$t_{ij} \rightarrow \frac{t}{\sqrt{q}}$ to obtain a non-trivial model
\cite{vollmetz}.
The lattice hamiltonian is then mapped onto an
equivalent impurity problem which is
supplemented by a selfconsistency condition.
The detailed derivation of the resulting mean field equations
is already given in great detail elsewhere \cite{review,gk,gkq}.
We shall assume a semi-circular bare
density of states
$\rho^o (\epsilon) =
(2 / {\pi D}) \sqrt{1 - ( \epsilon /D)^2}$, with $t=\frac{D}{2}$,
which is 
realized in a Bethe lattice
and on a fully connected fully frustrated version of
the model \cite{zrk,agwk,rkz}.
This density of states 
has the same edge behavior as one obtained from
a three dimensional tight binding model.
As there is only one $3d-$electron in the $Ca_xSr_{1-x}VO_3$ 
system, we set $\mu = 0$ that renders the model half-filled.
We use the  second order perturbative calculation (IPT)
to iteratively solve the associated impurity problem \cite{zrk}.
It has already been shown elsewhere that this method is in excellent
agreement with the exact solutions that can be obtained using
exact diagonalization or quantum Monte Carlo techniques \cite{review}.
Still, we use the IPT
calculation since the exact methods either lack the resolution
or cannot reach the low temperature limit that is relevant
for the present discussion.

The solution of the Hubbard model in the limit of large dimensions 
has been obtained recently \cite{motthubb}. 
One of the basic results is the existence
of a MIT at an intermediate value of the interaction
$U_c \approx 3D$ ($U_c = 3.37D$ whithin IPT).
As the transition is approached by increasing $U/D$,
one finds that the density of states splits into a characteristic
three-peak structure. There is a
central quasiparticle peak  at the Fermi
energy of width $\epsilon_F^*$
that narrows as the transition is approached indicating 
the enhancement of the effective mass.
This feature is accompanied by the incoherent lower
and upper Hubbard bands that emerge at frequencies of the order of
$\pm \frac{U}{2}$ and which acquire the spectral weight that is
transferred from the lower frequencies as the central peak narrows
\cite{gk,zrk}.

From the theoretical standpoint, the $Ca_xSr_{1-x}VO_3$ compound is
extremely interesting, since the chemical substitution
of a $Sr^{2+}$ ion by $Ca^{2+}$ of the same
valence modifies the $V-O-V$ bond angle
from $\sim 180^o$ to $160^o$, thus controlling 
the strength of the ratio $U/D$.
Therefore for concentrations ranging from $x=0$ to $1$ we go from
a less correlated to a strongly correlated metal.

We now turn to the comparison between the experimental data and the theoretical
predictions of the model Hamiltonian. 
The strategy will be to extract the relevant model parameters, 
i.e. $U$ and $D$, from the experimental
data, and use them as sole input
for the theoretical model.
We emphasize that, following this procedure,
there are no free parameters left to adjust in the theoretical
fit of the experimental data.

\begin{figure}
\centerline{\epsfxsize=2.8truein
\epsffile{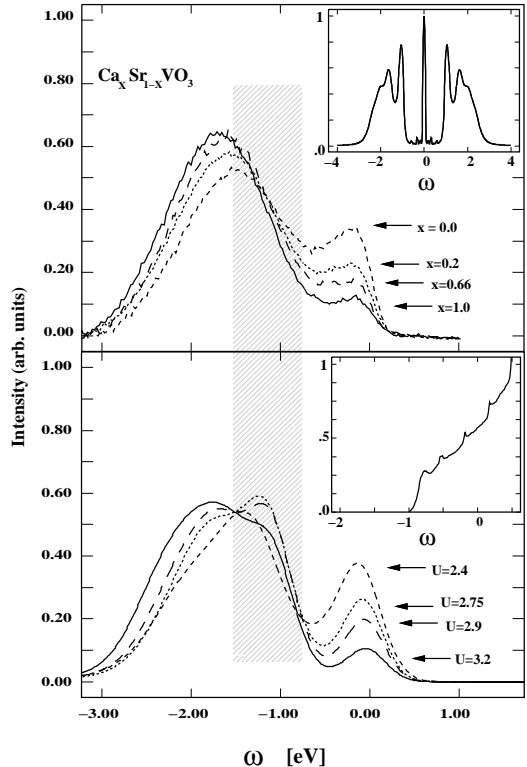}}
\caption{ Above: Experimental photoemission spectrum of $Ca_xSr_{1-x}VO_3$
at 80K (from \protect\cite{inoue} with background from oxygen $2p$
band subtracted).
The inset contains the theoretical density of states $\rho(\omega)$
obtained from IPT at $T=0$ for a value of the interaction
${U\over D} =3$ close to the MIT.
Below: Theoretical density of states $\rho(\omega)$ for the model parameter
$U$ indicated in the figure and $D=1eV$
obtained by IPT.
The data corresponds to the $\omega < 0$ part of $\rho(\omega)$ convoluted
with the experimental resolution.
The inset shows the corresponding
LDA results (from \protect\cite{inoue}).
}
\label{fig1}
\end{figure}

We first consider the recent
experimental photoemission data of Inoue {\it et al.} \cite{inoue} 
which we reproduce in Fig.\ref{fig1}.
The low energy photoemission
spectra consist of two contributions: a broad feature at higher frequencies
and a another one, much smaller and narrow, that 
occurs at the Fermi energy. 
As we increase the value of $x$, i.e., the correlation strength, we observe
a systematic transfer of spectral weight 
between the two spectral features \cite{inoue}. 
The lineshape of this spectrum is very reminiscent of the results for 
the density of states that are obtained from the large dimensional 
approach to the Hubbard model
close to the MIT point (see inset of Fig.\ref{fig1}).
Thus, we associate the higher frequency feature to the lower Hubbard
band (LHB) and the feature at the Fermi level to the 
low frequency quasiparticle 
contribution.
We extract the model parameters noting that:
{\it i)} the LHB features peak at approximately $-\frac{U}{2}$;
{\it ii)} the half-width of the LHB should approximately correspond
to $D$; and {\it iii)} the ratio of the quasiparticle peak to the LHB
spectral weight is controlled by the ratio $U/D$ in our model.

Photoemission experiments probe the occupied part of the density of
states. Accordingly, we have multiplied
the theoretical spectrum by the Fermi distribution function of $80K$,
and then convoluted
with a Gaussian function of width $0.3eV$  that corresponds to the finite
experimental resolution \cite{inoue}. 
In Fig.\ref{fig1} we display the results for the
theoretical photoemission intensity. 
The estimated values for $U$ and $D$ are shown in the figure.
We have check that a good fit can be also obtained by small variation
of the parameters. Although one expects that $D$ would also 
systematically change with composition, we choose to set equal to 
constant $1eV$. This choice not only provides a reasonable fit, but it further
allows to clearly appreciate the systematic evolution of the model lineshape
as a function of the interaction.

It is apparent that the theory can successfully capture 
the basic features of the experimental data. 
We observe the transfer of spectral weight between the LHB and the 
quasiparticle peak as a function of the interaction.
In particular, we note that the theory captures correctly the
intermediate frequency range around which the transfer of weight occurs
(indicated with a shaded region in the figure).
These results are more remarkable in regard 
to the complete
failure of the LDA method to capture even the most basic features
of the data (inset Fig.\ref{fig1}) \cite{inoue}.

Although we find that the basic features
of the transfer of spectral weight are very 
well captured by our model, upon
a more detailed comparison one notes that some
discrepancies remain. 
Especially in regard of the shape of the small
quasiparticle peak and the lower energy edge of the LHB.
The discrepancies in the shape of the quasiparticle contribution, 
in particular, the 
renormalization of the value of $\rho(\omega)$ at the Fermi level,
as was pointed out in \cite{inoue,morikawa},
may be due to a basic feature of the large-d approach, namely,
the locality of the self-energy. 
This is probably connected to the overestimation by a factor of $\sim 4$ of
the enhancement of the calculated mass renormalization 
compared with that deduced
from the measured electronic specific heat coefficient ($\gamma_{exp} 
\approx 9.2mJ
/K^2 mol$ for $CaVO_3$ \cite{Ishikawa}, this
value is about 3 times larger than the LDA prediction). 
This issue may possibly be resolved by a better treatment of the
spatial fluctuations.
Nevertheless, we emphasize 
that the goal of the present paper is not
the description of the lineshape in its full detail, but rather, the
understanding of some basic features in the distribution and transfer
of spectral weight by the effect of correlations.

\begin{figure}
\centerline{\epsfxsize=2.8truein
\epsffile{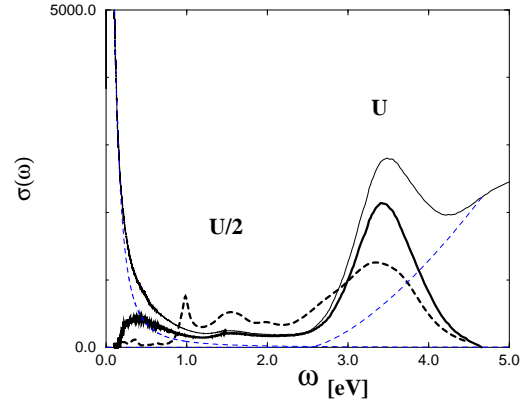}}
\caption{The experimental optical conductivity $\sigma(\omega)$ in units
of $[\Omega cm]^{-1}$ for $CaVO_3$
at $T=300K$ (thin line). The mid-infrared spectra (bold line)
obtained after the
subtraction of the background contribution ($\sim (\omega - 2.5eV)^{3/2}$)
and a lorentzian fit of Drude part at lower frequencies (thin dashed lines).
The model optical conductivity (bold dashed line)
for ${U\over D} =3.2$ and $D=1eV$
obtained by IPT (arbitrary units).
}
\label{fig2}
\end{figure}

We now turn to the comparison of our model results for the
optical conductivity \cite{jfp} with the new experimental data that
we report here. The optical reflectivity measurement was carried out for
$CaVO_3$ which is the most strongly correlated case.
A single crystal 
sample was grown by the floating zone method and specularly
polished. 
It was annealed in air at $\sim 470K$ for 24 hours in order to make
the oxygen concentration become stoichiometric \cite{inoue2}.
The reflectivity spectrum was measured at room temperature in energy range 
from $0.03eV$ to $6.5eV$, using a Fourier transform spectrometer
($0.03eV-0.5eV$) 
and a grating monochromator ($0.5eV-6.5eV$). To obtain the optical
conductivity by Kramers-Kronig analysis, we supplemented our data
with that of Kasuya {\it et al.} \cite{kasuya} obtained in a higher energy
region. Also a high frequency asymptotic
function of the form 
$\omega^{-4}$ was used, and we extrapolated
our data to the lowest frequencies  using the Hagen-Rubens formula.
The result is displayed in Fig.\ref{fig2}. It basically presents
three contributions: a low frequency narrow peak which is the Drude part
of the optical conductivity, a second small feature that appears at energies
around $1.75eV$ and a third large contribution that peaks at $3.5eV$.
The Drude part at low frequencies corresponds to
the metallic character of the compound. However, unlike in normal metals, 
the  Drude contribution is only a 
fraction of the total optical spectral weight.
This is a realization of the original arguments of Kohn for the optical 
response of systems close to a MIT \cite{kohn}.

We can readily, and rather intuitively, interpret the higher frequencies 
peak structure in terms of excitations between the three
features present
in the density of states ({\it cf.} upper inset of Fig.\ref{fig1}).  
The large contribution in the conductivity spectrum 
at $\omega \sim  3.5eV$ corresponds to the
direct excitations from the lower to the upper Hubbard bands, therefore
this contribution should peak at an energy of the order of the Hubbard
interaction $U$. Accordingly, the position 
of the peak very approximately corresponds to
twice the energy of the maximum of the lower Hubbard band in the photoemission
spectrum of Fig.\ref{fig1}. 
Also, we note that the large spectral weight of this feature corresponds to 
the fact that most of the single particle spectral weight is concentrated
in the incoherent Hubbard bands.
On the other hand, we should also expect
optical excitations from the LHB to
the quasiparticle peak and from this peak to the upper Hubbard band.
As the energy for these
two types of excitations is approximately $\frac{U}{2}$, they should also
contribute to the optical spectrum with a feature at frequencies
$\omega \sim \frac{U}{2} \approx 1.75eV$ from the estimation made for
this parameter in the photoemission discussion. 
This is in fact observed in the experimental
results. We can further understand the small spectral
weight of this feature from the fact that the quasiparticle
peak in the $\rho(\omega)$ has only a small fraction of the total spectral
intensity.

In Fig.\ref{fig2} we also present the optical spectrum obtained from the LISA
method using the model parameters determined before ($x=1$).
We find that all the basic features of our
previous qualitative discussion are actually realized 
in the solution of the model. 
For a clearer comparison of the model prediction to
the experimental data,
we have subtracted in the latter the high frequency background due
to excitations from the oxygen band 
to the UHB. 
{}From a fit to the lower energy tail of the oxygen band we estimate
this background contribution to start at 
$\omega \approx 2.5eV$.
Since our model does not contain disorder, the Drude part
is a $\delta-$function at the origin. Thus, also for the sake
of a clear comparison, we have subtracted the Drude part of the 
experimental spectra by fitting a lorentzian function at low frequencies.
The remaining part of the experimental spectra then corresponds to
the mid-infrared features at $\frac{U}{2}$ and $U$ that we discussed above.
As the results show, these features are well captured by our model
at the precise positions and with qualitatively correct relative weights.
As in the case of the photoemission data, the remaining differences
in the lineshapes may be possibly due to
limitations of the IPT or the election of a semicircular $\rho^0$.

A final quantity that is useful in the interpretation of 
optical conductivity data is the ratio of the Drude part of the 
optical response to the total spectral weight. In an uncorrelated
system, this ratio is one. The effect of the correlations that drive the
system close to the MIT is to split the contributions
to $\sigma(\omega)$ into a $\delta$-function like part at 
low frequency that corresponds to the coherent
quasiparticle conduction 
and a regular mid-infrared contribution describing the
incoherent motion. Within our present mean field theory, this ratio
is obtained as  $R=Z\frac{\langle K_o \rangle}{\langle K \rangle}$, where
$Z$ is the quasiparticle residue and $\langle K_o \rangle$ is the 
expectation vaue of the kinetic energy in the free system ($U=0$)
\cite{rkk,rkk2}. From the experiment we find that $R_{exp}\approx 0.3$, which
is in good agreement with the prediction of the theory $R=0.25 \pm 0.03$
(the uncertainty comes from the approximate determination of the
model parameters $U$ and $D$ and the rapid variation of $R$ close to the
MIT). 
We note that $1/R$ is usually called 
the optical mass renormalization and that in the $d\rightarrow \infty$ limit
$\frac{m^*}{m}=1/Z$. Thus, in our mean field theory 
these two quantities are exactly
related by the renormalization of the kinetic energy by the interaction.

An important observation is that the consistent picture that
emerges from our {\it simultaneous}
discussion of photoemission and ac-conductivity
gives further support to
the assumption that a single band model may be
used for the study of the $Ca_xSr_{1-x}VO_3$ system. 
This observation is rather surprising, as the typical magnitude
of crystal fields $\sim 0.1eV$ is unlikely to explain
how the 3-fold degeneracy of the $t_{2g}$-bands is effectively lifted.
A more plausible explanation may follow from the consideration
of a more realistic degenerate model with
inter-orbital exchange and correlation interaction. 
The solution of such a model
poses extra technical challenges and work along this line is currently
under way.

Finally, we would like to briefly mention that the 
temperature dependence of the dc-resistivity follows a
$\rho \sim AT^2$ behavior which demonstrates that the system is, as
our model solution,
a Fermi liquid.

The satisfactory agreement between the
theory and the experimental data demonstrates that the 
Hubbard model treated within a mean field theory
that becomes exact in 
the limit of large lattice coordination is indeed a good strating point 
for the study of electronic systems where the competition of localization
and itinerancy play a crucial role.
The LISA solution of the single band model 
allows us to understand the basic features
that are experimentally observed 
in the spectroscopy of $Ca_xSr_{1-x}VO_3$.
We find that the transfer of weight observed in photoemission
data and the unusual distribution of spectral weigh in the optical
response are related to
the proximity of the system to a MIT point 
characterized by the
emergence of the small energy scale $\epsilon_F^*$ \cite{rkz}.
Thus we obtain within {\it a single model with essentially no
adjustable parameters}, a consistent interpretation of the unusual
features observed in {\it both} experiments.

We acknowledge valuable discussions with Y. Tanaka, Y. Shimoi, 
S. Abe, G. Kotliar and A. Georges. This work was partially supported 
by the Agency of Industrial Science and Technology - Japan Industrial 
Technology Association (AIST-JITA) Invitation Program for Foreign 
Researchers.

\end{document}